\documentclass[twocolumn,showpacs,amsmath,amssymb]{revtex4}

\usepackage{graphicx}
\usepackage{dcolumn}
\usepackage{bm}

\newcommand{\bb}       {\ensuremath {\beta\beta}}
\newcommand{\tbb}      {\ensuremath {2\nu\beta\beta}}
\newcommand{\nbb}      {\ensuremath {0\nu\beta\beta}}
\newcommand{\gamgam}   {\ensuremath {\gamma\gamma}}
\newcommand{\nuc}[2]   {\ensuremath {^{#1}\mathrm{#2}}}
\newcommand{\gray}     {$\gamma$-ray}
\newcommand{\qbb}      {Q$_{\beta\beta}$}
\newcommand{\gamr}     {$\gamma$ ray}
\newcommand{\gamrs}    {$\gamma$ rays}

\begin{document}

\title{Double Beta Decay of $^{100}$Mo to Excited Final States}

\author{M. J. Hornish}
 \altaffiliation[Present address: ]{Department of Physics and Astronomy, 
Ohio University, Athens, OH 45701.}
 \email{hornish@ohio.edu}
\author{L. De Braeckeleer}
 \altaffiliation[Present address: ]{Departamento de F\'{\i}sica, 
Universidad de los Andes, Bogot\'a, Colombia.}
\affiliation{The Department of Physics, Duke University and Triangle
Universities Nuclear Laboratory,\\ Durham, NC  27708-0308}

\author{A. S. Barabash}
\author{V. I. Umatov}
\affiliation{Institute for Theoretical and Experimental Physics, \\
B.Cheremushkinskaya 25, 117259, Moscow, Russia}

\date{\today}

\begin{abstract}
A systematic study of the inclusive ($0\nu+2\nu$) double beta (\bb)
decay of \nuc{100}{Mo} to various excited final states of \nuc{100}{Ru} 
was performed.  Utilizing two large HPGe detectors operated in 
coincidence, a search for the subsequent deexcitation \gamgam\ 
cascades was conducted.  A 1.05-kg sample of isotopically enriched 
($98.4\%$) \nuc{100}{Mo} was investigated for 455 days, yielding an 
unambiguous observation of the \bb\ decay of \nuc{100}{Mo} to the 
$0_1^+$ state (1130.3 keV) of \nuc{100}{Ru}.  This excited final state 
decays via the $0_1^+ \rightarrow 2_1^+ \rightarrow 0_{gs}^+$ sequence 
(with $E_{\gamma1} = 590.8$ keV and $E_{\gamma2}=539.5$ keV), and 22 
such coincidence events were detected, with a continuous background 
estimated to be 2.5 events.  This counting rate corresponds 
to a decay half-life for the \bb($0^+\rightarrow0_1^+$) transition of 
$T_{1/2}^{(0\nu+2\nu)} = \left[6.0^{+1.9}_{-1.1} (\rm{stat}) \pm 
0.6(\rm{syst})\right] \times 10^{20}$ years.  Lower limits on decay 
half-lives were achieved for higher
excited final states.
\end{abstract}


\maketitle

\section{\label{sec:level1}Introduction}
A considerable escalation in the attention paid to the study of double
beta (\bb) decay is owed, in no small part, to the tremendous progress
made by recent neutrino oscillation experiments.  The
remarkable success of solar and atmospheric neutrino experiments 
(see reviews \cite{Hol02,Bah02,Mal03}) represents a profound achievement 
and a veritable breakthrough in our understanding of the neutrino.  
More recently, the KamLAND reactor antineutrino result \cite{Egu03} and 
the latest SNO result \cite{Ahm04} further reinforce the evidence for
the oscillation of neutrinos.  Indeed, there can 
no longer be any doubt that the neutrino is, in fact, a massive particle.  

At the same time, however, for all of the progress that has been 
accomplished in measuring the mass differences ($\Delta m^2$) for
neutrinos of unlike flavors, such oscillation experiments are inevitably
and inherently insensitive to two additional important properties: (1) 
the neutrino's absolute mass scale and (2) the Majorana-Dirac nature of 
the neutrino.  The only present experimental method capable of attempting  
to simultaneously answer both of these questions is the search for the 
neutrinoless mode (\nbb) of double beta decay.  Although this transition
has not yet been observed beyond a reasonable doubt, the successful 
detection of this decay mode may also provide useful information 
about the neutrino mass hierarchy (normal, inverted, or quasi-degenerate) 
and lepton-sector CP violation via a measurement of the Majorana 
CP-violating phases (see discussions \cite{Bil04,Pas03,Pas04}).

The recent surge in interest in \bb\ decay has initiated the proposal
of several large-mass experiments that will search for the neutrinoless
mode (see review \cite{Ell02,Ell04,Bar04b,Avi05}).  The interpretation of
any future results from these experiments depends on the reliable
estimation of nuclear matrix elements ($M^{0\nu}$) for the \nbb\ 
transition.  While the phase-space factor for this decay can be 
accurately calculated, any
uncertainty in the corresponding nuclear matrix elements will adversely 
affect the ability to correctly extract Majorana neutrino mass 
information.  Several theoretical models have been devised, including
the well-known quasiparticle random phase approximation (QRPA), but a 
thorough analysis of the accuracy of the various models to properly
calculate the nuclear matrix elements is difficult at present, although
efforts are being made (see, e.g., \cite{Suh05,Rod05}).

One testing ground for our understanding of the \bb-decay 
nuclear matrix elements is through the study of the ordinary allowed
second-order weak decay, namely the two-neutrino double beta (\tbb) 
decay.  In general, expanding our experimental knowledge of the \tbb\ 
process for a variety of different nuclei will improve the
overall comprehension of the nuclear part of double beta decay.
Although the nuclear matrix elements for this
mode ($M^{2\nu}$) are not identical to those of the neutrinoless mode,
experimental measurements of $M^{2\nu}$ can be compared to theoretical
model calculations as a check of the validity of the various 
theoretical schemes upon which the models rely.

Over the past decade and a half, significant progress has been made in
successfully detecting the \tbb\ decay to ground states for several
nuclei, including \nuc{48}{Ca}, \nuc{76}{Ge}, \nuc{100}{Mo} and
\nuc{150}{Nd} (see summary \cite{Bar02a}).  However, for certain nuclei, 
there are low-lying excited levels in the daughter nuclide that may also 
be populated by \bb\ decay.  Although these excited-state transitions 
suffer from a reduced phase space, and thus a longer half-life, it is 
important to realize that such decays present an additional quantity to 
measure, namely the deexcitation \gamrs.  In fact, in the framework of 
QRPA models, excited-state transitions exhibit a much different 
dependence on the particle-particle strength parameter $g_{pp}$ than 
ground-state transitions \cite{Gri92,Aun96,Suh98}.  Hence, different 
aspects of the theoretical techniques used to calculate nuclear matrix 
elements may be probed by the study of decays to excited states.

For the case of \nuc{100}{Mo}, the Q values for the \bb\ decay to some of 
the excited states of \nuc{100}{Ru} are large enough that the detection
of the \tbb\ decay may be possible.  The idea \cite{Bar90} to search
for the \gamrs\ emitted from the deexcitation of the excited
daughter nucleus prompted several measurements, which used
enriched molybdenum samples, to operate \gray\ detectors in singles
mode in underground, very-low background settings.  By detecting the
subsequent gamma decay of the excited daughter nuclide rather than
the two beta particles, this type of measurement is not sensitive to 
the particular mode of decay.  However, the emitted \gamrs\ from
the excited nucleus have a fixed energy, so if they can be detected with
good energy resolution and relatively high efficiency, the measurement
is rather straightforward and the relevant experimental background can
be substantially suppressed.

\begin{figure}
  \includegraphics[width=3.0in]{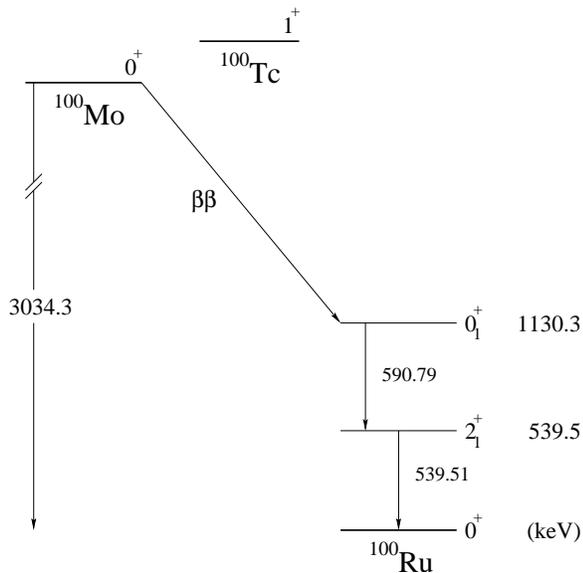}
  \caption{\label{fig:moscheme}Simplified \bb-decay scheme 
of \nuc{100}{Mo} to the $0_1^+$ excited state, whose deexcitation
proceeds 100$\%$ of the time through the intermediate $2_1^+$ state.}
\end{figure}
In particular, the study of \nuc{100}{Mo} focused primarily on the 
\tbb\ decay to the $0_1^+$ state of \nuc{100}{Ru} (1130.3 keV) by
searching for the subsequent $0_1^+ \rightarrow 2_1^+ \rightarrow 
0_{g.s.}^+$ decay cascade, with $E_{\gamma1} = 590.79$ keV and 
$E_{\gamma2} = 539.51$ keV (see Figure \ref{fig:moscheme}).
One such experiment was performed by measuring enriched \nuc{100}{Mo}
samples ($\sim1$ kg) with a low-background HPGe detector.  An analysis
of the single \gray\ spectrum positively identified the \tbb $(0^+ 
\rightarrow 0_1^+)$ transition, and a deduced half-life of $T_{1/2} = 
(6.1^{+1.8}_{-1.1}) \times 10^{20}$ years was produced \cite{Bar95}.
Afterwards, a second experiment using the same technique
yielded a measured half-life for this transition of $T_{1/2} = 
(9.3^{+2.8}_{-1.7}) \times 10^{20}$ years, with a systematic error 
estimated to be $\sim$ 15\% \cite{Bar99}.  By summing the single \gray\ 
spectrum of Ref. \cite{Bar95} to that of Ref. \cite{Bar99}, one obtains 
a combined half-life of $T_{1/2} = (7.6^{+1.8}_{-1.1}) \times 10^{20}$ 
years, also with a systematic error of 15\% \cite{Bar99}.  On a related 
note, it is interesting to point out that a recent experiment using a 
similar technique successfully observed the \tbb\ decay of \nuc{150}{Nd} 
to the $0_1^+$ state of \nuc{150}{Sm}, with a half-life estimated to be 
$T_{1/2} = \left[1.4^{+0.4}_{-0.2} (\rm{stat}) \pm 0.3(\rm{syst})\right] 
\times 10^{20}$ years \cite{Bar04a}.

The preliminary results from the present experiment on \nuc{100}{Mo}, 
with a quoted half-life of this \tbb $(0^+ \rightarrow 0_1^+)$
transition of $T_{1/2} = \left[5.9^{+1.7}_{-1.1} 
(\rm{stat}) \pm 0.6 (\rm{syst})\right] \times 10^{20}$ years, 
were previously presented \cite{DeB01} and will be explored in greater
detail in this Article.

\section{Experimental Details}
By relying on single \gray\ searches, previous measurements of the 
\tbb($0^+ \rightarrow 0_1^+$) decay of \nuc{100}{Mo} have involved the
implementation of low-background detection systems, where detectors
were made from low-radioactivity materials.  These experiments were 
operated in underground laboratories, where the natural overhead
shielding reduces cosmic-ray induced backgrounds.  These experiments 
produce single \gray\ 
spectra exhibiting an excess of counts in the regions of interest near 
539.5 keV and 590.8 keV, above a rather substantial background.  In the 
two cases mentioned in Section \ref{sec:level1}, the corresponding 
signal-to-background ratios were approximately 1:7 \cite{Bar95} and 
1:4 \cite{Bar99}.  Extracting the total number of counts attributed to 
the \bb\ decay depends greatly on a reliable fit to the background.

The technique utilized in the present Article involves the use of 
two low-background HPGe detectors operated in a coincidence scheme, the
first time such a \gamgam\ coincidence technique has been applied to 
the study of \bb\ decay.  In some cases, as in the present one, the 
reduced detection efficiency of the coincidence measurement can be more 
than compensated by a corresponding suppression of the associated 
background.  This alternative approach to background reduction relies on 
the simultaneous detection of the two \gamrs\ emitted from the 
deexcitation of the daughter nuclide.  Owing to the unique \gray\ 
energies, the coincidence background can be suppressed to such an extent 
that the signal-to-background ratio may be much higher than in single 
\gray\ searches, even in above-ground experiments.

\subsection{TUNL-ITEP Apparatus}
Conducted in the Low Background Counting Facility of the Triangle
Universities Nuclear Laboratory (TUNL), the experimental work is 
centered around a \gamgam\  coincidence apparatus, the cornerstone of 
which is the operation in coincidence of two high-purity germanium (HPGe) 
detectors.  These large, custom-made HPGe detectors (p-type coaxial) were 
specially designed to have a large frontal surface area.  The size of each 
HPGe crystal was 8.8 cm in diameter and 5.0 cm in thickness.  Each 
detector was fabricated from low-background materials, and each crystal 
was coupled to a very-low background cryostat in the J-type configuration, 
which is used not only to isolate the detector crystal from the 
preamplifier, the HV filter and LN$_2$ dewar, but also to allow for its 
easy insertion into an anticoincidence annulus.  The two HPGe detectors 
have 1.8-keV FWHM energy resolution at 1.33 MeV and 0.8-keV resolution at 
0.122 MeV, with the relevant peak-to-Compton ratio rated as 80.4.  

\begin{figure}
\includegraphics[width=3.5in]{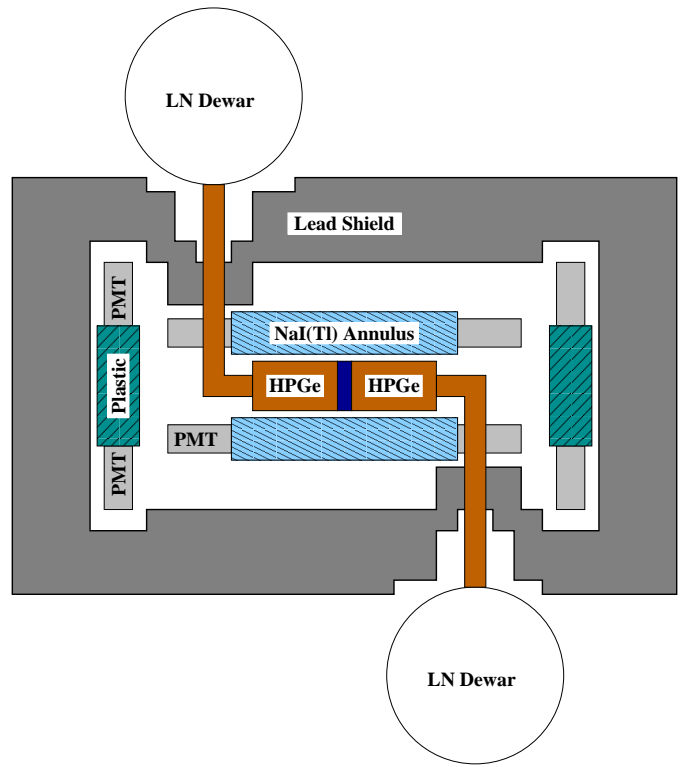}
\caption{\label{fig:apparatus}(Color online) Schematic view (not to scale) 
of the TUNL-ITEP \bb-decay apparatus.  The enriched ($98.4\%$) disk of 
\nuc{100}{Mo} is sandwiched between the two HPGe detectors.}
\end{figure}

To suppress the inherent background contributions that result from 
primary cosmic-ray radiation, secondary cosmogenically-induced radiation 
and primordial radiation, active and passive shielding techniques were 
employed.  The active shielding consists of anticoincidence counters, 
which include two plastic plate scintillators (10-cm thickness) and one 
NaI(Tl) annulus (56-cm length, 35.6-cm diameter with a 12.5-cm hole along
the axis of symmetry) into which the source and HPGe detectors are 
inserted.  These counters also veto potential backgrounds arising from
unwanted Compton scattering events within the HPGe detectors.  The passive 
shielding includes a lead-brick enclosure and the overhead shielding 
afforded by the room ceiling and the floors of the building ($\sim$10 m 
w.e.).  A schematic view of the apparatus can be seen in Figure 
\ref{fig:apparatus}. 

A reliable energy calibration of the HPGe detectors was attained by 
monitoring known background $\gamma$ peaks [238.6 keV (\nuc{212}{Pb}), 
511.0 keV (annihilation), 609.3 keV (\nuc{214}{Bi}), and 1460.8 keV 
(\nuc{40}{K})].  Various performance checks were used to verify the 
proper operation
of the detectors and the electronics over the course of the experiment.
For example, daily monitoring of the location of the more prominent
background $\gamma$ peaks allowed for any corrections due to gain shifts.
Secondly, the sensitivity of the veto system was readily checked
by monitoring the strength of known single gamma lines (e.g., 1461-keV
\gamrs\ from \nuc{40}{K}), which in principle are not vetoed, against 
those emitted in coincidence with additional gamma quanta (e.g., 
511-keV annihilation \gamrs), which will normally be vetoed.

For the study of the \tbb\ decay of \nuc{100}{Mo} to excited
states of \nuc{100}{Ru}, an enriched metallic disk was fabricated and 
loaned to TUNL by Russian collaborators from the Institute of Theoretical
and Experimental Physics (ITEP).  It was enriched in 
\nuc{100}{Mo} to 98.4$\%$ and had a mass of 1.05 kg.  The disk was 
approximately 1.1 centimeters thick and 10.6 
centimeters in diameter, which more or less matches the size of the two 
HPGe detectors.  The disk thickness
was chosen in order to achieve a suitable balance between maximizing the 
source mass while limiting efficiency losses due to attenuation and 
geometry effects.  Preceding the study of this enriched disk, a metallic 
disk of natural molybdenum (1-kg mass, 10-cm diameter, 0.965-cm 
thickness, 9.6$\%$ \nuc{100}{Mo}) was studied to better understand 
potential backgrounds in the coincidence spectra.

\subsection{Efficiency Determination}
In order to properly interpret the \gamgam\ coincidence spectra
obtained, a solid understanding of the efficiency of the apparatus
to detect \gamgam\ coincidence events must be achieved.  For the purpose 
of the present study, it was realized that a direct measurement of the 
efficiency was not only feasible, but it would also provide the most 
reliable information.  Using a source of known strength, the principle 
behind the efficiency measurement is to observe a deexcitation cascade 
involving two \gamrs\ with energies and angular distributions comparable
to those of interest from the \bb-decay measurements.  As such, the 
radioactive \nuc{102}{Rh} nuclide was judiciously selected for this 
measurement.

\begin{figure}
\includegraphics[width=3.0in]{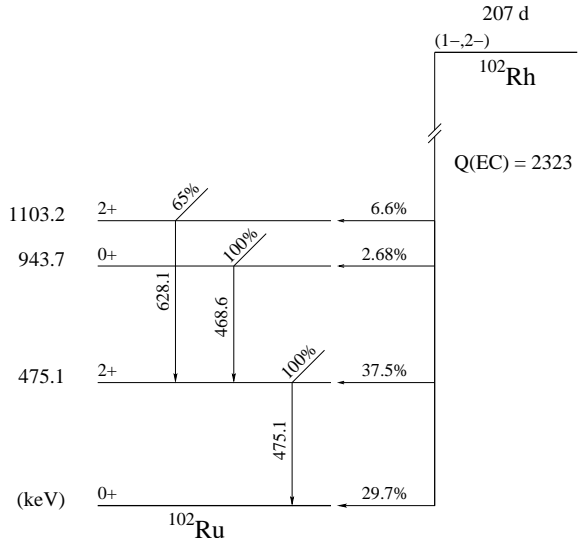}
\caption{\label{fig:rh102}Decay scheme of \nuc{102}{Rh} and the relevant 
gamma cascades of the excited states of \nuc{102}{Ru}.}
\end{figure}

The choice of the \nuc{102}{Rh} isotope is quite practical for many
reasons.  First, the half-life of \nuc{102}{Rh} is approximately 207 
days, which is convenient because the short-lived radioactivities 
decay away leaving a rather pure sample.  Moreover, the source can be 
used over long periods of time without having to apply significant
corrections to a set of measurements performed over a period of a few 
weeks.  Secondly, as shown in Figure \ref{fig:rh102},
following the electron capture of \nuc{102}{Rh} to the $0_1^+$ state of 
\nuc{102}{Ru} with a branching ratio of 2.68$\%$, the isotope emits two 
\gamrs\ in coincidence with energies (468.6 keV and 475.1 keV) similar 
to those from the $\beta\beta(0^+\rightarrow0_1^+)$ decay of 
\nuc{100}{Mo}.  Since $\beta^+$ decay does not occur, there is no 
radiation from bremsstrahlung or annihilation, thus making the 
measurement very simple.  At the same time, the probability to reach the 
$0_1^+$ state via electron capture to higher excited states is small 
($<8.8\%$), hence most of these pairs of \gamrs\ (468.6 - 475.1 keV) 
are emitted without any additional quanta that could interfere with the 
measurement by a summation effect.

\begin{figure}
\includegraphics[width=3.0in]{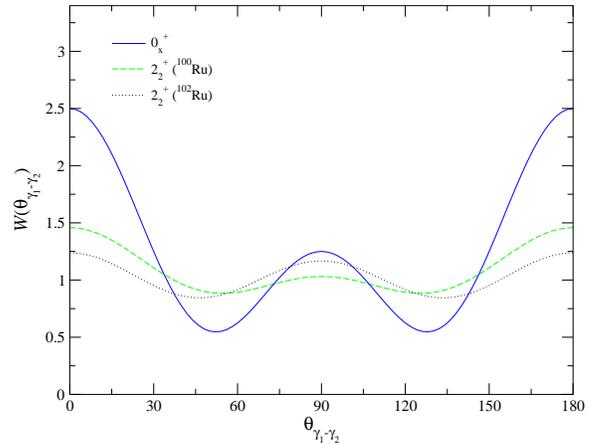}
\caption{\label{fig:angcor}(Color online) Calculated angular distribution, 
$W$, as a function of the angle $\theta_{\gamma_1-\gamma_2}$ between the 
two \gamrs\ emitted in the \gamgam\ cascade.  The solid curve is the 
angular correlation for all $0^+ \rightarrow 2^+ \rightarrow 0^+$ decay 
sequences.  The dashed (dotted) curve is the correlation between the two 
\gamrs\ emitted from the $2_2^+ \rightarrow 2_1^+ \rightarrow 0^+$ 
sequence of \nuc{100}{Ru} (\nuc{102}{Ru}).}
\end{figure}

In addition, the angular correlation of the two emitted \gamrs\ is 
identical to those of the \gamgam\ cascades that follow the \bb\ decay
of \nuc{100}{Mo} to excited $0^+$ final states because it proceeds via 
the $0^+ \rightarrow 2^+ \rightarrow 0^+$ sequence.  This condition
arises because the corresponding deexcitation cascade proceeds solely 
through electric quadrupole (E2) radiation, with no mixing of other 
multipolarities.  Finally, one 
additional advantage of \nuc{102}{Rh} as the source is the presence of 
a higher excited $2^+$ state at 1103.2 keV of \nuc{102}{Ru}, 
corresponding to a subsequent $2^+ \rightarrow 2^+\rightarrow0^+$ 
cascade that provides a measure of the detection efficiency for such a 
deexcitation sequence.  Hence, the efficiency for detecting \bb\ 
decays to excited $2^+$ states can also be determined, albeit in a more
complicated manner because the mixing of multipolarities (M1 and E2) 
depends on the specific \gamr\ involved (see Figure \ref{fig:angcor}).

A \nuc{102}{Rh} source was produced at TUNL by bombarding a target made 
from natural ruthenium (31.6$\%$ \nuc{102}{Ru}) with a 5-MeV proton 
beam, thereby generating \nuc{102}{Rh} via (p,n) activation of \nuc{102}{Ru}.
Following the production of the \nuc{102}{Rh} source, a small portion of
it (0.2cm $\times$ 0.2cm $\times$ 0.1cm) was isolated for use in the 
efficiency measurement.  Its strength was determined by comparing the 
intensity of the 475.1-keV gamma line to that of the 661.6-keV gamma line 
from a calibrated \nuc{137}{Cs} source.  To reproduce the photon 
attenuation in molybdenum that occurs in the \bb-decay experiment, the 
source was placed between thin natural molybdenum disks.  Ten disks (10-cm 
diameter and 0.1-cm thickness) were used to ensure that the overall 
thickness was the same as the original experiment, thereby creating 
the same spacing between the two HPGe detectors.  The use of these disks 
also enabled the variation of the position of the \nuc{102}{Rh} source 
with respect to the molybdenum disks, as a function of cylindrical 
coordinates (radius, depth, and azimuthal angle) with respect to the
center of the detectors.

\begin{figure}
\includegraphics[width=3.0in]{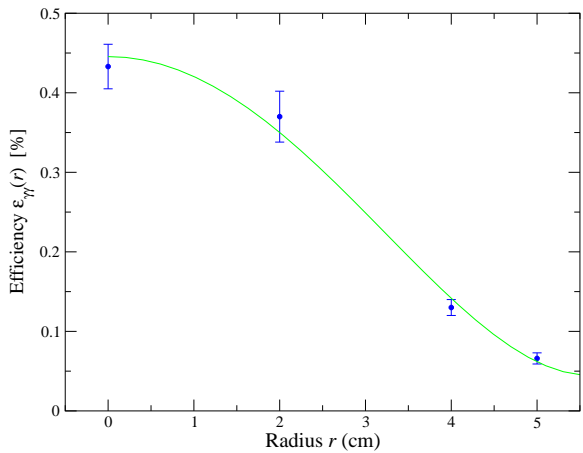}
\caption{\label{fig:effrad}(Color online) The efficiency 
$\varepsilon_{\gamma\gamma}(r)$ to detect 591 - 540 keV coincidences, 
measured as a function of the radial distance $r$ from the center of the 
detectors, and an appropriate fit to the data points.}
\end{figure}

At each location of the \nuc{102}{Rh} source, the number of \gamgam\
coincidence events involving the 469 - 475 keV cascade, arising from
the decay of \nuc{102}{Rh} to the $0_1^+$ state of \nuc{102}{Ru}, were 
counted.  The results were compared to the total number of decays that
actually occurred, as given by the source strength and the branching
ratio.  Measurements of the \gamrs\ emitted from the \nuc{102}{Rh} 
source were adjusted to corresponding \nuc{100}{Ru} $\gamma$-decay
sequences by applying corrections for both the relative efficiency of
the HPGe detectors and the photon attenuation in Molybdenum.
Figure \ref{fig:effrad} displays the results for the radial dependence
of the efficiency $\varepsilon_{\gamma\gamma}(r)$ to detect the 591 - 
540 keV \gamgam\ coincidence events from the \nuc{100}{Mo} \bb\ decay.
[The notation used henceforth, such as 591 - 540 keV coincidence, refers 
to \gamgam\ coincidence events in which a count in one detector 
is observed in the energy interval $591\pm2.5$ keV while at the same time
a count is registered in the second detector in the interval $540\pm2.5$ 
keV.  The notation applies to all possible \gamgam\ combinations
and energies, and the energy interval used ($\pm2.5$ keV) is reflective
of the detector resolution.]  

An integration over the \nuc{100}{Mo} source volume yields the total 
detection efficiency $\varepsilon_{\gamma\gamma}^{tot}$, and the results 
for the \gamgam\  cascades of interest from excited states in 
\nuc{100}{Ru} are presented in Table \ref{tab:mototeff} for the 
enriched \nuc{100}{Mo} disk.
\begin{table*}
\caption{\label{tab:mototeff}Calculations of the efficiencies 
$\varepsilon_{\gamma\gamma}^{tot}$ ($\%$) to detect \gamgam\  cascades 
resulting from the \bb\ decay of \nuc{100}{Mo} to various excited states 
of \nuc{100}{Ru}.}
\begin{ruledtabular}
\begin{tabular}{ccccc}
Transition  &  Level (keV) &  $\gamma_1 - \gamma_2$ (keV)  &  Branching Ratio $f_b$  &  \mbox{$\varepsilon_{\gamma\gamma}^{tot}$ ($\%$)}  \\
\hline
$0^+\rightarrow0_1^+$ & 1130.3 & 590.8 - 539.5   & 1.00 & $0.219\pm0.022$  \\ 
$0^+\rightarrow2_2^+$ & 1362.2 & 822.4 - 539.5   & 0.58 & $0.196\pm0.020$  \\  
$0^+\rightarrow0_2^+$ & 1741.0 & 1201.4 - 539.5  & 0.59 & $0.185\pm0.019$  \\
		      &        & 378.9 - 1362.0  & 0.17 & $0.196\pm0.020$  \\  
$0^+\rightarrow0_3^+$ & 2051.5 & 1512.1 - 539.5  & 0.86 & $0.165\pm0.017$  \\
	 	      &        & 689.4 - 1362.0  & 0.06 & $0.161\pm0.016$  \\ 
$0^+\rightarrow0_4^+$ & 2387.4 & 1847.8 - 539.5  & 0.51 & $0.142\pm0.014$  \\
	 	      &        & 1025.2 - 1362.0 & 0.21 & $0.154\pm0.015$  \\  
\end{tabular}
\end{ruledtabular}
\end{table*}
A Monte Carlo simulation was also performed to check the validity of 
these values.  It incorporated all known factors, including the full-energy
peak efficiency of the HPGe detectors, the strongly anisotropic angular
correlation of the \gamrs\ and their attenuation in the sample, and effects
of the extended geometry.

\section{Analysis}

\subsection{Background Considerations}
Before proceeding with an analysis of the data from the TUNL-ITEP 
apparatus, it is important to address potential background 
contributions that may arise in the \gamgam\ coincidence spectrum.
Although the discussion that follows pertains to the \bb\ 
decay of \nuc{100}{Mo} to the $0_1^+$ state of \nuc{100}{Ru}, the 
principle arguments can be easily extended to decays involving higher
excited states (i.e., different \gray\ energies).

Due to the excellent energy resolution afforded by the HPGe detectors, 
the primary concern is background processes that can generate 
an identical gamma decay cascade with no additional quanta.  Based on 
the present criteria to detect excited-state double beta 
decays by means of the simultaneous detection of the two \gamrs\ 
subsequently emitted via deexcitation, any such background source would 
in principle be indistinguishable from the desired experimental 
signature.  

Realizing that the cascade involving two, and only two, \gamrs\ with 
energies of 591 keV and 540 keV is unique to the deexcitation of the 
$0_1^+$ state of \nuc{100}{Ru}, the major background consideration is 
of any mechanisms other than \bb\ decay that may populate this excited 
level.  Taking into account the fact that the \gamgam\ detection 
efficiency rapidly decreases as one moves away from the center of the 
\nuc{100}{Mo} disk, as seen in Figure \ref{fig:effrad}, background 
sources that are of the greatest concern are those originating from 
within the source itself.  

As such, the most relevant background candidate would appear to be a 
significant flux through the detector setup of cosmic-induced protons, 
which can in principle result in the ($p$,$n$) activation of \nuc{100}{Mo} 
to bound states in \nuc{100}{Tc}.  The subsequent $\beta$ decay of this 
intermediate nucleus is delayed by the 15.8-s half-life, thereby 
rendering the anticoincidence counters ineffective in vetoing this 
process.  The branching ratio of the \nuc{100}{Tc} $\beta$ decay to 
the $0_1^+$ state of \nuc{100}{Ru} is 5.7$\%$.  Unfortunately, most of 
the remainder ($93\%$) of \nuc{100}{Tc} $\beta$ decays proceed to the 
ground state, which prevents using the \gamgam\ coincidence spectrum to 
search for additional gamma decay cascades that might otherwise provide 
a measure of the ($p$,$n$) activation of \nuc{100}{Mo}.

An alternative means of estimating the proton flux in the apparatus and 
hence estimating the likelihood of background contributions from 
\nuc{100}{Mo}($p$,$n$) is to search for the very similar ($p$,$n$) activation
of \nuc{95}{Mo} to the isomer ($1/2^-$, 61-day half-life) of \nuc{95}{Tc}.
The electron capture of $^{95\mathrm{m}}$Tc populates two levels (1039.2
keV and 786.2 keV) of  \nuc{95}{Mo} with large branching ratios ($30.1\%$
and $38.0\%$, respectively).  The 1039.2-keV state decays to the ground
state $89\%$ of the time via a cascade of two \gamrs\ (835.1 keV and 
204.1 keV), while the 786.2-keV state proceeds to the ground state
with a \gamgam\ cascade (where E$_{\gamma1}$ = 582.1 keV and 
E$_{\gamma2}$ = 204.1 keV) $78\%$ of the time.  The intermediate 204-keV
level has a negligibly short half-life of 0.75 ns.  Searching for the
835 - 204 keV and 582 - 204 keV coincidence events provides a convenient
method for measuring the proton flux in the laboratory.

To accomplish this study, a metallic disk of natural molybdenum 
(1-kg mass, 10-cm diameter, 0.965-cm thickness, $15.9\%$ \nuc{95}{Mo}) 
was investigated in the TUNL-ITEP apparatus for 180 days.  Analysis of
the data revealed exactly zero 835 - 204 keV coincidence events.  For
the 582 - 204 keV coincidence, there was no excess of counts observed 
beyond a fit to the background.  Combining these two results allows one 
to gauge the potential for any contribution of ($p$,$n$) activation in the 
\nuc{100}{Mo} sample, assuming a similar proton flux.  The conclusion 
from the analysis of the \nuc{95}{Mo} study is that ($p$,$n$) activation 
contributes less than one count per year to the important 591 - 540 keV
coincidence events of interest.

Another potential background candidate involves sources arising from a
significant neutron flux through the detectors and the \nuc{100}{Mo} 
disk.  There exist two known mechanisms that could, in principle, 
provide unwanted \gamgam\ coincidence events that would, in practice, be 
indistinguishable from the \bb\ decay to excited final states in the
present experiment.  

First, if a ruthenium impurity resides in or near the \nuc{100}{Mo} disk, 
elastic scattering of neutrons on \nuc{100}{Ru} (12.6$\%$ abundance) 
could produce a 591 - 540 keV coincidence that at present would
register as a spurious \nuc{100}{Mo} \bb-decay event to the $0_1^+$ 
state of \nuc{100}{Ru}.  A measure of the significance of this possible
background can be obtained by searching for similar neutron scattering
processes, such as \nuc{100}{Mo}($n$,$n'\gamma$) and
\nuc{76}{Ge}($n$,$n'\gamma$), where the abundance of \nuc{76}{Ge} is 
$7.44\%$.  The former can result in an excited state of \nuc{100}{Mo} 
that decays to the ground state through a \gamgam\  cascade with 
$E_{\gamma1}=528.2$ keV and $E_{\gamma2}=535.6$ keV, whereas the latter 
produces a \gamgam\  cascade with $E_{\gamma1}=545.5$ keV and 
$E_{\gamma2}=562.9$ keV from the deexcitation of the \nuc{76}{Ge}
excited state.

During the experimental trial that studied the enriched \nuc{100}{Mo} 
disk, approximately 12 events per year fit the 528 - 536 keV coincidence 
profile.  Also, for the neutron scattering process on the \nuc{76}{Ge} 
nucleus, the 546 - 563 keV coincidence was observed approximately four times 
per year.  Making realistic assumptions about differences in cross
section, detection efficiency, and a reasonable estimate on an upper limit
of the size of such an impurity within the \nuc{100}{Mo} disk,
a conservative estimate of the threat of this 
neutron scattering background amounts to no more than approximately one 
half of one 591 - 540 keV coincidence event per year.

Secondly, there is a realistic, albeit unlikely, alternative scenario 
that could result in a coincidence 
involving 591-keV and 540-keV \gamrs\ that do not originate from 
the deexcitation of the $0_1^+$ state of \nuc{100}{Ru}.  Rather, 
following the neutron capture by a \nuc{100}{Mo} nucleus, 
\nuc{100}{Mo}($n$,$\gamma$), the $\beta$ decay of \nuc{101}{Mo} can 
populate excited states of \nuc{101}{Tc}, from which two deexcitation 
gamma rays ($E_{\gamma1}=540.1$ keV and $E_{\gamma2}=590.9$ keV) are 
emitted as part of a four-photon cascade.

In principle, summation and veto effects should rule out the likelihood
of any significant contribution from this background candidate.  
Furthermore, the wide, neutron-broadened peaks that result from the 
neutron scattering interactions on germanium nuclei are not accompanied 
by narrow peaks usually ascribed to thermal neutron capture.  Thus, the 
thermal neutron flux is most likely very small.  Even so, there is a 
small, non-zero probability that the two bystander photons could escape 
undetected, leaving behind a 591-keV photon in one HPGe detector and a 
540-keV photon in the other.  Conveniently, an analysis of this 
possibility is readily accomplished as a byproduct of the normal data 
compilation.

The transition to the 2218-keV level of \nuc{101}{Tc} constitutes only 
0.4$\%$ of \nuc{101}{Mo} $\beta$ decays, and is followed 0.69$\%$ of the 
time by two possible four-photon cascades containing both the 540.1-keV 
and the 590.9-keV \gamrs.  A more likely scenario is the $\beta$-decay to 
the 1319.6-keV level with a branching ratio of 6.8$\%$, which produces a 
\gamgam\  cascade of two photons with energies of 713.0 keV and 590.9 
keV.  This \gamgam\  cascade is approximately 1000 times 
as likely to occur as the other four-photon cascades.  An analysis of the 
455 days of data from the \nuc{100}{Mo} run reveals four such 713 - 591 keV 
coincidence events.  Thus, one can place a conservative upper limit of 
0.003 counts per year originating from 540 - 591 keV background events from 
the four-photon deexcitation cascades involved in the $\beta$ decay of 
\nuc{101}{Mo}.  Again, the actual contribution should, however, be
considerably less because the two other photons from the four-photon
cascades would normally lead to the veto of the signal.

Finally, for reference sake, it was found that truly random coincidences 
contribute less than 0.0003 counts yr$^{-1}$ for the \nuc{100}{Mo} 
experiment (with energies near 600 keV).  Moreover, this background 
should be relatively uniform over the region of interest and would not 
be specific to the gamma decay cascades relevant to the present study.

\subsection{\label{sec:main}Decay to the $0_1^+$ Excited State}
\begin{figure}
\includegraphics[width=3.0in]{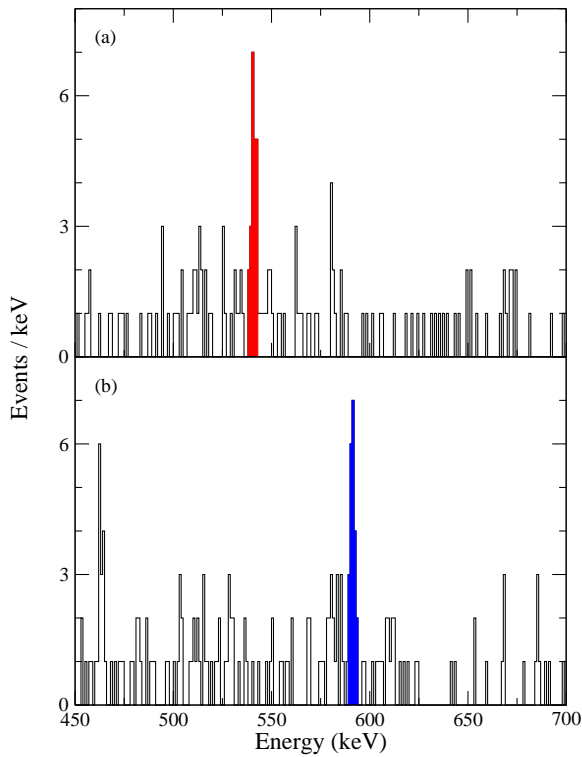}
\caption{\label{fig:datmo}(Color online) The \gray\  spectra in 
coincidence with $590.8\pm2.5$ keV (a) and with $539.5\pm2.5$ keV (b).  
The 22 observed 591 - 540 keV events (shaded) from the 
\bb$(0^+\rightarrow0_1^+)$ decay of \nuc{100}{Mo} were obtained in 455 
days of measuring time.}
\end{figure}

The search for the \bb\  decay of \nuc{100}{Mo} to the $0_1^+$ state of 
\nuc{100}{Ru} involves the simultaneous detection of the coincident 
deexcitation ($0_1^+ \rightarrow 2_1^+\rightarrow 0_{gs}^+$) 
\gamrs, where $E_{\gamma1} = 590.8$ keV and $E_{\gamma2} = 539.5$ keV 
(see Figure \ref{fig:moscheme}).  
This \gamgam\  cascade from the $0_1^+$ state of \nuc{100}{Ru} has a 
branching ratio of 100$\%$.  The isotopically-enriched (98.4$\%$) 
\nuc{100}{Mo} disk was measured for a period of 455 days in the TUNL-ITEP 
apparatus.  An inspection of the \gray\  spectra in coincidence with 
$539.5\pm2.5$ keV and $590.8\pm2.5$ keV, respectively (as shown in Figure
\ref{fig:datmo}), reveals 22 such 591 - 540 keV coincidence events.  The 
width of the coincidence window is five channels (5 keV) in order 
to account for the detector resolution.  The presence of the peaks in 
their respective spectra is unmistakable when comparing their amplitudes 
to that of the surrounding background.
A conservative measure of this background incorporates 
the full spectrum within $\pm100$ keV of the 591 - 540 keV coincidence 
peak, a measurement of which yields $2.50\pm0.25$ background counts per 
5-keV bin.

Deriving an empirical measure of the half-life of the excited-state 
\bb\ decay from the observed counting rate is straightforward.  
The 22 detected events ($\sigma_N=\sqrt{22} = 4.7$) yield 
$N_{\gamma\gamma} = 19.5 \pm 4.7$ events after subtracting the 
background, thereby corresponding to a signal-to-background ratio of 
approximately 8:1.  From the sample size (1.05 kg), the counting time 
($\sim1.25$ yr), and the detection efficiency (0.219\%), the calculated 
decay half-life for the present measurement is
\begin{equation}
	T_{1/2}^{(0\nu+2\nu)} \; = \; \left[6.0_{-1.1}^{+1.9}(\rm{stat})\pm0.6(\rm{syst})\right]\times10^{20} \; \rm{yr}.
	\label{eq:halflife}
\end{equation}
The asymmetrical statistical uncertainty results from the low number of
counts, while the systematic error entails the $10\%$ uncertainty in the 
detection efficiency.  It should be noted that this value for the
half-life differs slightly from the previously published value of 
$\left[5.9_{-1.1}^{+1.7}(\rm{stat}) \pm 0.6 (\rm{syst})\right] \times 
10^{20}$ years \cite{DeB01} for this experiment.  The difference arises 
from a slight modification to the calculated efficiency, the inclusion of 
some additional data (455 days of counting versus the original 440 days), 
and a corrected calculation of the statistical error bars.

From the decay half-life presented here, it is possible to extract a 
value for the nuclear matrix element corresponding to this \tbb\ 
transition.  Using  the phase-space factor $G = 1.64 \times 10^{-19}$ 
yr$^{-1}$ (for the axial-factor coupling constant $g_A = 1.254$), one 
obtains the nuclear matrix element $M_{2\nu}(0_1^+) = 0.101\pm0.013$, as 
scaled by the electron mass.

The detection scheme is incapable of deciphering the two-neutrino mode 
from the neutrinoless mode of \bb\ decay, thus the half-life of Equation
\ref{eq:halflife} is listed as an inclusive ($0\nu + 2\nu$) quantity.  
Nevertheless, using values for \nuc{100}{Mo} \bb\ transitions to the 
ground state of \nuc{100}{Ru} ($T_{1/2}^{2\nu}=6.75 \times 10^{18}$ yr 
\cite{Sil97} and $T_{1/2}^{0\nu}>5.5\times10^{22}$ yr, 90\% C.L. 
\cite{Eji01}), one concludes that the neutrinoless mode constitutes less 
than $0.012\%$ of \bb\ transitions to the ground state.  In fact, the 
most recent \nuc{100}{Mo} data from the NEMO3 experiment \cite{Arn04} 
restrict the neutrinoless \bb\ decay of \nuc{100}{Mo} to the ground state 
further: $T_{1/2}^{2\nu} = [7.68 \pm 0.02 (\rm{stat}) \pm 0.54 
(\rm{syst})] \times 10^{18}$ yr and $T_{1/2}^{0\nu} > 3.1\times10^{23}$ 
yr (90\% C.L.).  Thus, assuming a similar proportion for the 
excited-state \bb\ decays, it is reasonable to assume a negligible 
contribution from $0\nu$ events in the present data.  This conclusion is
further reinforced by theoretical predictions for the neutrinoless \bb\ 
decay of \nuc{100}{Mo} to the $0_1^+$ excited state, with the half-life
for this \nbb($0^+\rightarrow0_1^+$) transition estimated to be: 
$2.59 \times 10^{26}$ yr \cite{Suh02} and $(0.76 - 1.46) \times
10^{25}$ yr \cite{Sim01a} for $\langle m_\nu \rangle = 1$ eV.

\begin{figure*}
\includegraphics[width=6.0in]{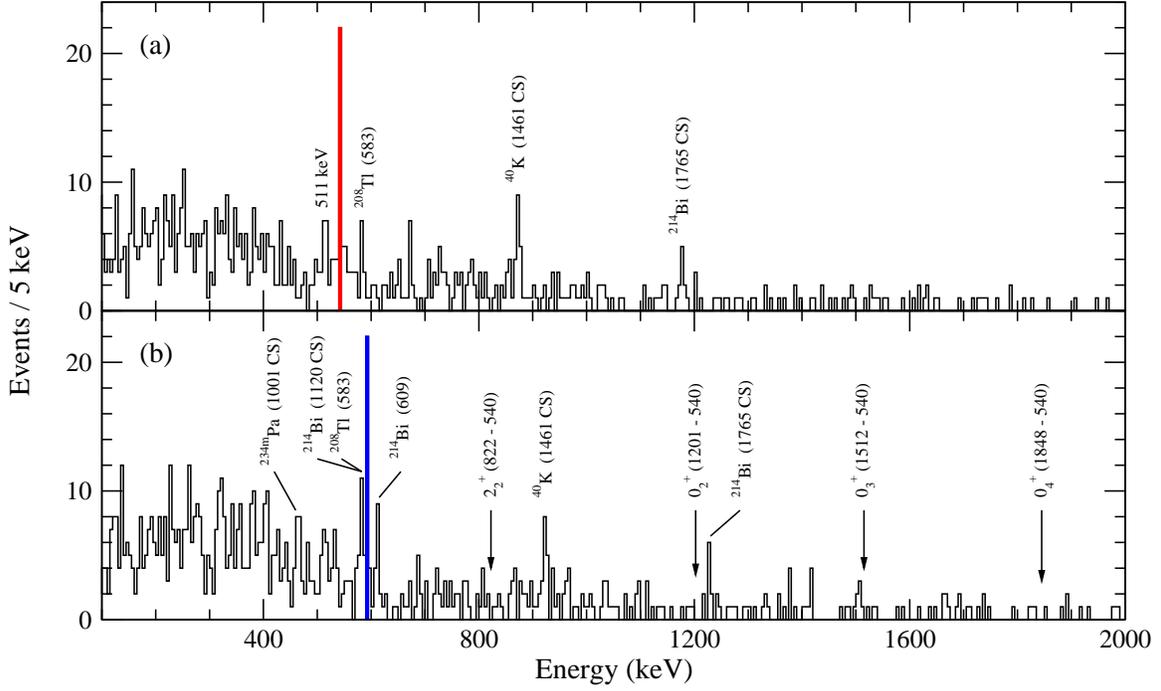}
\caption{\label{fig:datmo-5}(Color online) The \nuc{100}{Mo} \gamgam\  
coincidence data, compressed to include 5 keV per bin.  Shown are the 
\gray\  spectra in coincidence with $590.8\pm2.5$ keV (a) and with 
$539.5\pm2.5$ keV (b), with some of the prominent background lines 
identified (CS = Compton Scattering). Note the 22 observed 591 - 540 keV 
coincidence events from the \bb ($0^+\rightarrow0_1^+$) transition.}
\end{figure*}
In Figure \ref{fig:datmo-5}, the \gray\ spectra in coincidence 
with $539.5\pm2.5$ keV and $590.8\pm2.5$ keV show a broader range of 
energies and reveal some interesting features of the background.  Both
spectra exhibit the presence of some known gamma peaks, such as the 
annihilation peak (511 keV) and peaks from \nuc{208}{Tl} (583 keV) and 
\nuc{214}{Bi} (609 keV).  As expected, the background at lower energies
($<500$ keV) is generally more prominent.  This phenomenon is a 
consequence of two factors: (1) the larger photopeak detection efficiency
at lower energies and (2) the contributions at low energies from Compton
scattering events.  In the latter case, the Compton continua of a large 
number of the more prominent primordial background \gamrs, in addition
to those of interest in the present study, lie below 500 keV.

Another interesting feature that was observed in the \gamgam\ coincidence
spectra at higher energies ($>800$ keV) is the Compton scattering of at
least two known primordial background peaks.  The most noticeable instance
involved the 1461-keV \gamr\ (\nuc{40}{K}), which is always emitted in 
single with no accompanying gamma quanta.  The Compton scattering of this
photon can lead to the deposition of 591 (540) keV in one of the HPGe 
detectors and the remaining 870 (921) keV in the other HPGe detector, and
there are indeed rather prominent peaks at the respective energies of the
corresponding spectra of Figure \ref{fig:datmo-5}.  The same phenomenon
occurs for the 1764.5-keV \gamr\ (\nuc{214}{Bi}), producing the observed
1225 - 540 keV and 1174 - 591 keV coincidence events.  To a lesser extent,
there also appear to be some Compton scattering coincidence events 
corresponding to 1001.0-keV (\nuc{234m}{Pa}) and 1120.3-keV 
(\nuc{214}{Bi}) \gamrs.

\subsection{Decay to Higher Excited States}
\begin{figure}
\includegraphics[width=3.0in]{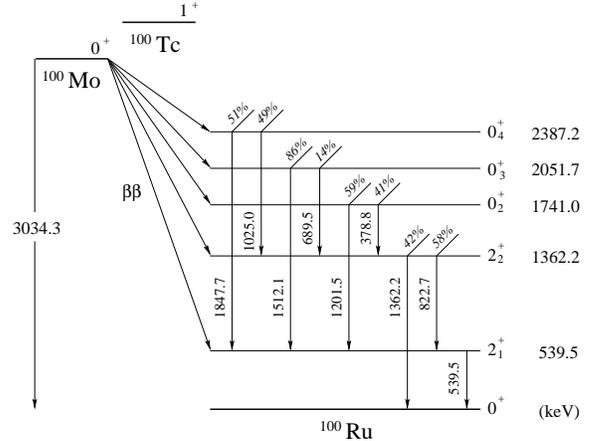}
\caption{\label{fig:moschemefull}Decay scheme for the \bb\ decay of 
\nuc{100}{Mo} to higher excited states of \nuc{100}{Ru}, excluding the 
$0_1^+$ state.  Shown are the subsequent gamma cascades and the 
corresponding \gray\ energies and branching ratios.}
\end{figure}

In addition to the transition to the excited $0_1^+$ state, the \bb\ 
decay of \nuc{100}{Mo} can also populate higher excited states of 
\nuc{100}{Ru}.  In principle, the measurement of these additional 
transitions is readily accomplished with the TUNL-ITEP apparatus by 
searching for alternative \gamgam\ coincidence events associated with 
the deexcitation of these higher excited states.  In the present study, 
the relevant \nuc{100}{Ru} levels of interest and the corresponding gamma 
decay cascades are shown in Figure \ref{fig:moschemefull}.  

Since these other processes were not observed in the present experiment, 
it is useful to briefly mention the procedure for setting a lower limit 
on a decay half-life.  In general, the half-life limit is given by
\begin{equation}
	T_{1/2} \; > \; \frac{(\ln 2) \, N_0 \, t \, f_b \, \varepsilon^{tot}_{\gamma\gamma}}{N_d},
	\label{eq:halflimit}
\end{equation}
where $f_b$ denotes the branching ratio of a given \gamgam\  cascade, and 
all relevant values for the detection efficiency 
$\varepsilon^{tot}_{\gamma\gamma}$ are taken from Table 
\ref{tab:mototeff}.  As recommended by the Particle Data Group 
\cite{PDBook}, the factor $N_d$, which represents an upper limit on the 
number of detected events above background, is determined by a 
statistical estimator derived for a process that obeys Poisson statistics.  

The deexcitation of the higher excited $0^+$ states of \nuc{100}{Ru} at 
1741.0, 2051.7, and 2387.2 keV can each occur via two different \gamgam\ 
cascades, one involving the transition through the $2_1^+$ state (539.5 
keV) and the other through the $2_2^+$ state (1362.0 keV).  Hence, the 
search for \bb\ decay to these excited states involves monitoring the 
\gray\ spectrum in coincidence with these two \gray\ energies.  It 
should be noted that, in addition to these 
two-photon cascades, there is also a three-photon deexcitation sequence.  
However, the existing apparatus is not sensitive to cascades involving
more than two \gamrs, thus such decay sequences will not be 
considered here.  At the same time, a search for the \bb\ decay to the
$2_2^+$ excited state is accomplished by searching for coincidence 
events involving 822.4-keV and 539.5-keV \gamrs, without the emission
of any additional gamma quanta.

An analysis of the 455 days worth of data yielded the following
results for the \bb\ decay of \nuc{100}{Mo} to higher excited states.  
For the transition to the $0_2^+$ state (1741.0 keV) of \nuc{100}{Ru},
zero 1201 - 540 keV coincidence events were detected and one 
379 - 1362 keV event was observed, which corresponds to a counting rate 
that is consistent with the background.  In the case of the transition 
to the $0_3^+$ state (2051.7 keV), a search revealed one 1512 - 540 keV 
event (again consistent with the corresponding background rate) and zero 
689 - 1362 keV events.  For the transition to the $0_4^+$ state (2387.2 
keV), zero 1848 - 540 keV events and zero 1025 - 1362 keV events were 
observed.  Lastly, a search for 822 - 540 keV coincidence events, 
corresponding to the \bb\ decay to the $2_2^+$ state, yielded zero 
counts.  Taking into consideration the associated \gamgam\ background 
rate in the regions of interest, limits on the \bb -decay transitions 
to higher excited states were calculated from Equation 
\ref{eq:halflimit} and are listed in Table \ref{tab:mohalflives}.
\begin{table*}
\caption{\label{tab:mohalflives}A summary of the experimental half-life values, or limits, for the \bb\ decay of \nuc{100}{Mo} to various excited states of \nuc{100}{Ru} corresponding to a 455-day counting period.  Limits are indicated at the 68(90)$\%$ C.L.  References for previous results are given in square brackets.}
\begin{ruledtabular}
\begin{tabular}{ccccc}
Transition  &  Level (keV)  &  \qbb\ (keV) & $T_{1/2}^{(0\nu+2\nu)}$ ($\times 10^{20}$ yr)  &  Previous results ($\times 10^{20}$ yr)  \\ 
\hline
$0^+\rightarrow0_1^+$ & 1130.3 & 1904.0 & $6.0_{-1.1}^{+1.9}(\rm{stat})\pm0.6(\rm{syst})$  &  $6.1_{-1.1}^{+1.8}$  \cite{Bar95}  \\
 &  &  &  &    $9.3_{-1.7}^{+2.8}$ \cite{Bar99}  \\
$0^+\rightarrow2_2^+$ & 1362.2 & 1672.1 & $>54(27)$  &  $>(13)$  \cite{Bar95} \\  
$0^+\rightarrow0_2^+$ & 1741.0 & 1293.3 & $>55(28)$  &  $>(13)$  \cite{Bar95} \\
$0^+\rightarrow0_3^+$ & 2051.7 & 982.6  & $>42(24)$  \\
$0^+\rightarrow0_4^+$ & 2387.2 & 647.1  & $>49(25)$  \\  
\end{tabular}
\end{ruledtabular}
\end{table*}

\section{Conclusions}
The implementation of a \gamgam\ coincidence counting technique
has resulted in a half-life of the \bb($0^+\rightarrow0_1^+$) transition
of \nuc{100}{Mo} of $\left[6.0^{+1.9}_{-1.1} (\rm{stat}) \pm 0.6 
(\rm{syst})\right] \times 10^{20}$ yr, and the corresponding nuclear 
matrix element is calculated to be $M_{2\nu}(0_1^+) = 0.101 \pm 0.013$, as 
scaled by the electron mass.  This half-life is in very good agreement 
with previous results for this transition \cite{Bar95,Bar99}, and 
represents an independent confirmation of this decay.  Moreover, the 
success of the present experiment and the methods used is exemplified by 
a superior signal-to-background ratio.  Although the decay to 
additional excited final states was not observed in the present 
measurement, stricter limits on the decay half-lives for these 
transitions were extracted.

This result is particularly important in the context of the single state
dominance hypothesis (SSDH) \cite{Aba84,Civ98,Civ99}.  Using the result 
for the EC transition \nuc{100}{Tc} $\rightarrow$ \nuc{100}{Ru} 
\cite{Gar93}, a prediction for the \bb\ decay of \nuc{100}{Mo} to the 
$0_1^+$ state of \nuc{100}{Ru} has been made: $T_{1/2}^{2\nu} = 4.45
\times 10^{20}$ yr \cite{Sim01b}.  Unfortunately, this prediction
has an accuracy of only 50\% due to the uncertainty of the measured
EC transition.  However, if in the future there is significant 
improvement in the accuracy of the EC transition, it may be interesting
to compare the experimental half-life as presented here to the
theoretical prediction to check on the validity of the SSDH.

On a more general note, it is interesting to point out that the present 
\gamgam\ counting technique represents a powerful tool that could 
potentially be extended to future, large-scale neutrinoless double beta 
decay experiments, such as Majorana \cite{Aal02}, GENIUS \cite{Kla01}, 
SuperNEMO \cite{Bar02b}, and CUORE \cite{Arn02}, for example.  Although 
\nbb\ transitions to excited final states 
have smaller \qbb\ values, the emission of fixed-energy, deexcitation
\gamrs\ opens the possibility of detecting the \nbb\ decay by way of a
multiple, unambiguous coincidence involving the successful detection of 
both the emitted $\beta$ particles as well as the \gamrs.  Such a multiple
coincidence technique would result in a dramatically-reduced background,
provided the decay products are detected with high efficiency and with 
good energy resolution.  

Indeed, the information that can be gained from searching for 
neutrinoless double beta decay to excited final states is just as 
valuable as that from the transition to the ground state.  Furthermore, 
depending on the particular experimental design and apparatus, it may in 
principle be possible to incorporate the option to search for \nbb\ 
transitions to excited final states, in conjunction with the search for 
the transition to the ground state, with minimal cost or reconfiguration.  
Adding this possibility would be considerably advantageous as it would 
serve to greatly augment the breadth of a given experiment, thereby 
increasing the potential for discovery.

\begin{acknowledgments}
The authors would like to thank Werner Tornow, Christopher Gould, and 
Frank Avignone, III for their help and advice on the development and 
execution of this project.  This work was supported in part by the U.S. 
Department of Energy, Office of High Energy and Nuclear Physics, under 
Grant No. DE-FG02-97ER41033.  
\end{acknowledgments}

\newpage 
\bibliography{mjhldb}

\end{document}